\documentstyle[aps,epsfig]{revtex}

\input{epsf.sty}
\def\framegraphics{\def\ifframe{\iftrue}}
\def\dontframegraphics{\def\ifframe{\iffalse}}
\def\drawgraphics{\def\ifdraw{\iftrue}}
\def\dontdrawgraphics{\def\ifdraw{\iffalse}}

\dontframegraphics
\drawgraphics

\newcommand{\graphics}[6]{
\def\epsfsize##1##2{#6##1}
\begin{picture}(#2,#3)
  \ifframe
    \put(0,0){\framebox(#2,#3){}}
  \fi
  \ifdraw
    \put(0,#3){\begin{picture}(0,0)
                 \put(#4,#5){\epsfbox{#1}}
               \end{picture}}
  \fi
\end{picture}}

\begin{document}

\preprint{
\vbox{
\hbox{July 2000}
\hbox{...}
}}

\title{Deconfinement in QCD at Finite Temperature}

\author{F.M.~Steffens}

\address{	Instituto de F\'{\i}sica -- USP,
		C.P. 66 318, 05315-970,
		S\~ao Paulo, Brazil}

\maketitle

\begin{abstract}

We present a simple and intuitive picture for the deconfinement of 
quarks and gluons at finite temperature: as the temperature increases,
QCD behaves like QED at $T=0$. We show this by calculating the QCD 
coupling constant as a function of the temperature and of the 
external momenta used to probe quarks and gluons. 

\end{abstract}

\vspace*{1cm}

The study of QCD at finite temperature and density is of crucial 
importance for the understanding of hadron formation in the early 
universe: as the temperature decreased, quarks
and gluons must have experimented a phase transition from a quark-gluon 
dominated universe to a hadron dominated universe.
Efforts to recreate the quark $\leftrightarrow$ hadron transition are
currently being fully implemented at RHIC, where it is hoped that
clear signals of a quark-gluon plasma phase will be detected.

It is also currently believed, thanks to lattice simulations
of finite temperature QCD \cite{book}, that a deconfined phase is 
expected to exist at temperatures above (and around) 0.2 $GeV$. 
However, we still
do not have a clear picture of how QCD itself changes from a 
confining to a non-confining theory. That is, the quarks which were originally
enslaved inside a region of $\sim 1\; fm$ are now allowed to propagate
over a larger portion of space, outside the region occupied by
the original nucleon. This is the 
strict sense that the word deconfinement is used througout this work. 
On the other hand, the difference between a non-confining 
theory, QED in the case, and a confining theory, QCD, resides exactly
in the fact that the later has a gauge self coupling\footnote{
QED at strong coupling can also have a confining 
phase \cite{kondo}. However, in our case, we will be considering only the
region of small external momentum where, as usual, 
$\alpha_s^{QED} \sim 1/137$.}. 
Physically, it may help to think that the deconfinement proccess
considered here involves a transition from a non Abelian to an Abelian
theory.  In this sense,
we think that in a deconfined phase, QCD should be ``less'' 
non Abelian than in a confined phase.

It is well known that at $T=0$, besides a color factor,
the QED and 
QCD $\beta$ functions would be the same if only fermion loops 
existed. The difference in sign between the QED and QCD $\beta$ functions
comes, essentially, from the gluon and ghost loops in the gluon
self energy (plus, of course, the vertex corrections). Alternatively,
we notice that the two beta functions are the same in the limit
that the number of colors, $N_c$, goes to zero. 
At finite temperature, it would be natural to expect
that the same kind of limit exists. 
We will see in this letter that, in fact, QCD at high temperature and small 
external momenta squared behaves like QED at $T=0$,
a conclusion naturally drawn from the QCD $\beta$ function
at finite temperature. However, the calculation 
of this quantity has been plagued with dificulties, most of them
related to the lack of gauge invariance \cite{kobes,chaichian,sasaki} 
and the 
dependence of the renormalized coupling at finite temperature 
on the vertex chosen to renormalize it \cite{fujimoto,nakkagawa,baier}.

To ensure the gauge invariance of the $\beta$ function
means to ensure the gauge invariance of the
vertex and self energy functions. Our freedom to choose among 
the different vertices (quark-gluon, gluon-gluon, ghost-gluon) 
should not affect the calculation of the running coupling.
In general, we have:

\begin{equation}
\alpha_s = Z^{-1}_\alpha \alpha_B,
\label{e5}
\end{equation}
where $Z_\alpha$ is the renormalization constant of the coupling.
If the gluon-ghost vertex is used for the renormalization, then:

\begin{equation}
Z_\alpha = \frac{\tilde Z_{T1}}{\tilde Z_{T3} Z_{T3YM}},
\label{e6}
\end{equation}
with $\tilde Z_{T1}$, $\tilde Z_{T3}$, and $Z_{T3YM}$, the gluon-ghost vertex,
the ghost propagator, and the gluon propagator renormalization constants,
respectively, calculated at finite temperature. 
Alternatively, we could use the
triple gluon vertex:

\begin{equation}
Z_\alpha = \frac{Z_{T1YM}}{Z^2_{T3YM}}.
\label{e7}
\end{equation}
Here, $Z_{T1YM}$ is the renormalization constant of the triple gluon
vertex. A similar expression exists for the quark vertex.

In a consistent calculation, $Z_\alpha$ is independent of the vertex 
used - at zero temperature this is a well known 
consequence of the Slavnov-Taylor identities. At finite temperature
the situation is the same, as long as we restrain ourselves
to the dominant, gauge invariant, terms as given by the 
Hard Thermal Loop (HTL) resummation program of Braeten and 
Pisarsky \cite{braaten}.
In fact, within this program one can show that the dominant 
terms obey the Abelian Ward identities 
\cite{braaten,taylor}:
 
\begin{equation}
k^\mu \Pi_{\mu\nu}(k) = 0,
\label{e8}
\end{equation}

\begin{equation}
k^\mu \Gamma_{\mu\nu\rho}(k,p,q) = \Pi_{\nu\rho}(q) - \Pi_{\nu\rho}(p).
\label{e9}
\end{equation}
The HTL approximation consists of disregarding the {\it external} momenta in
the numerators of the loop integrals, as the main contribution to these
integrals comes from the region where the {\it internal} momenta are of 
order of $T$, with $T$ taken to be large. 
Hence, if we compare the gluon, quark and ghost self energies, we
see that the later is subleading to the formers because it does not
have enough powers of internal momenta in the numerator to produce
the leading ($T^2$) behavior in the temperature.
From Eq. (\ref{e9}) we see that the ghost vertex will be 
subleading with respect to the quark and gluon vertices as well.
Taking this into account, it follows from Eqs. (\ref{e6}) and (\ref{e7})
that the thermal parts of the triple gluon vertex renormalization
constant and of the gluon self energy renormalizion constant should
be equal:

\begin{equation}
Z_{T1YM}^{\left(leading \; T^2\right)} = Z_{T3YM}^{\left(leading \; T^2\right)}.
\label{e10}
\end{equation}
This relation is valid only for the dominant, gauge invariant, 
contribution. $Z_\alpha$ could also be calculated through the
quark vertex. In this case there would appear an other constraint for
the leading $T^2$ terms. Instead of Eq. (\ref{e10}), we would 
have that the renormalization constants for the quark 
vertex and the quark self energy should be the same.

A direct computation of the transverse part of the gluon 
self energy gives \cite{braaten2,brandt}:

\begin{equation}
\Pi_T^{(1)} = - \left(N_c + \frac{n_f}{2}\right) \frac{g^2 T^2}{12 |\vec k|^2}
\left[\frac{k_0}{|\vec k|}
ln\left(\frac{k_0 + |\vec k|}{k_0 - |\vec k|}\right) - 
2 \frac{k_0^2}{k^2}\right],
\label{e1}
\end{equation}
where $k$ is the four momentum of the external gluon.
Only the dominant term in the high temperature 
expansion was written because of its gauge invariance. One can see
that Eq. (\ref{e1}) vanishs for $k_0 << |\vec k|$, and that it is
reduced to

\begin{equation}
\Pi_T^{(1)} = \frac{\alpha_s}{\pi} \left(N_c + \frac{n_f}{2}\right) 
\frac{2 \pi^2}{3} \frac{T^2}{\mu_T^2}
\label{e2}
\end{equation}
for $k_0 >> |\vec k|$, where $\alpha_s \equiv g^2/4\pi$ and 
$\mu_T$ is a mass scale ($\mu_T^2 \equiv |\vec k|k_0 $).
Including the $T=0$ part, we can calculate the renormalization constant
for $\Pi_T$, which will depend on $T/\mu$:

\begin{equation}
\Pi_T^R = Z_{T3YM} \Pi_T,
\label{e3}
\end{equation}

\begin{eqnarray}
Z_{T3YM} &=& 1 + Z_{T3YM}^{(1)} \nonumber \\ 
&=& 1 -  \frac{\alpha_s}{\pi} \left(N_c + \frac{n_f}{2}\right) 
\frac{2 \pi^2}{3} \frac{T^2}{\mu^2} 
- \frac{\alpha_s}{\pi}\left[N_c \left(\frac{13}{24} - \frac{a}{8}\right) 
- \frac{1}{6}n_f\right]\left(\frac{1}{\epsilon} + ln \mu^2 \right),
\label{e4}
\end{eqnarray}
where $\mu$ is the renormalization scale, $a$ is the gauge parameter,
and $\epsilon$ comes from the 
dimensional regularization of the $T=0$ part of the gluon
self energy \cite{pascual}. The $a$ dependence in Eq. (\ref{e4})
is cancelled by the remaining renormalization constants in Eqs. (\ref{e6}) 
and (\ref{e7}).
Finally, we notice that the calculation of the triple gluon vertex gives 
a $T^2/\mu_T^2$ dependence
for the dominant term in the same form as Eq. (\ref{e2}), in a way that
we can define a thermal renormalization constant which always satisfies
Eq. (\ref{e10}).

The calculation of the thermal $\beta$ function (including the $T=0$ and
the $T \neq 0$ parts), for a fixed temperature but an arbitrary 
renormalization point is now straightforward. Using Eqs. (\ref{e5}), 
(\ref{e6}), and (\ref{e4}), we have:

\begin{equation}
\mu\frac{d\alpha_s}{d\mu} = 
\frac{\alpha_s^2}{\pi}\left[-\frac{11}{6}N_c + \frac{2}{6} n_f 
- \mu \frac{dB(T^2,\mu^2)}{d\mu}\left(N_c + \frac{n_f}{2}\right)  \right],
\label{e11}
\end{equation}
to order $\alpha_s^2$, where $B(T^2, \mu^2) \equiv 2/3 \; \pi^2 \; T^2/\mu^2$.
This is the only Renormalization Group Equation for $\alpha_s$ because
there is only one renormalization scale for both the $T=0$ and $T \neq 0$
parts of the renormalized gluon self energy.
The solution of Eq. (\ref{e11}) is:

\begin{equation}
\alpha_s(Q^2) = \frac{\alpha_s(Q_0^2)}
{1 + \frac{\alpha_s(Q_0^2)}{4\pi}\left[\left(\frac{11}{3}N_c 
- \frac{2}{3}n_f \right)ln \left(\frac{Q^2}{Q_0^2}\right) 
+ 4 (B(T^2,Q^2) - B(T^2, Q_0^2)) \left(N_c + \frac{n_f}{2}\right)\right]}.
\label{e12}
\end{equation}
It is helpful to rewrite Eq. (\ref{e12}) in the same format of the
$T=0$ theory. To this end, we define an effective number of colors
and an effective number of flavors:

\begin{equation}
N_c^{eff} = \left[1 + \frac{12}{11}\frac{B(T^2,Q^2) - B(T^2, Q_0^2)}
{ln \left(\frac{Q^2}{Q_0^2}\right)} \right] N_c,
\label{e13}
\end{equation}

\begin{equation}
n_f^{eff} = \left[1 - 3 \frac{B(T^2,Q^2) - B(T^2, Q_0^2)}
{ln \left(\frac{Q^2}{Q_0^2}\right)} \right] n_f.
\label{e14}
\end{equation}
With the help of Eqs. (\ref{e13}) and (\ref{e14}), the expression
for the running coupling is written as:

\begin{equation}
\alpha_s(Q^2) = \frac{\alpha_s(Q_0^2)}
{1 + \frac{\alpha_s(Q_0^2)}{4\pi}\left[\frac{11}{3}N_c^{eff} - 
\frac{2}{3}n_f^{eff}\right] ln \left(\frac{Q^2}{Q_0^2}\right)}.
\label{e15}
\end{equation}

The interesting point of this set of equations is that for 
fixed $Q_0^2$ and $T^2$, $N_c^{eff} < N_c$ and $n_f^{eff} > n_f$ 
as $Q^2 \rightarrow 0$. It implies that at some small value of the
external momenta squared, the coupling between quarks and gluons becomes
that of an Abelian theory.
To quantify this assertion, we make an explicit calculation of the coupling
as a function of $Q^2$ for some fixed values of $T$.
For $\alpha_s(Q_0^2)$, we use the
experimental value measured at $m_Z \approx 91 \; GeV$ 
and at zero temperature \cite{pdg}. 
That is, we assume that at such high values of the virtuality of the probe, 
temperatures of the order of $1 \; GeV$ are not relevant, an approximation
which amounts to set 
$B(T^2 = 1 \; GeV^2, Q_0^2 \approx 83 \times 10^3 \; GeV^2) \approx 0$.

\begin{figure}[htb]
\graphics{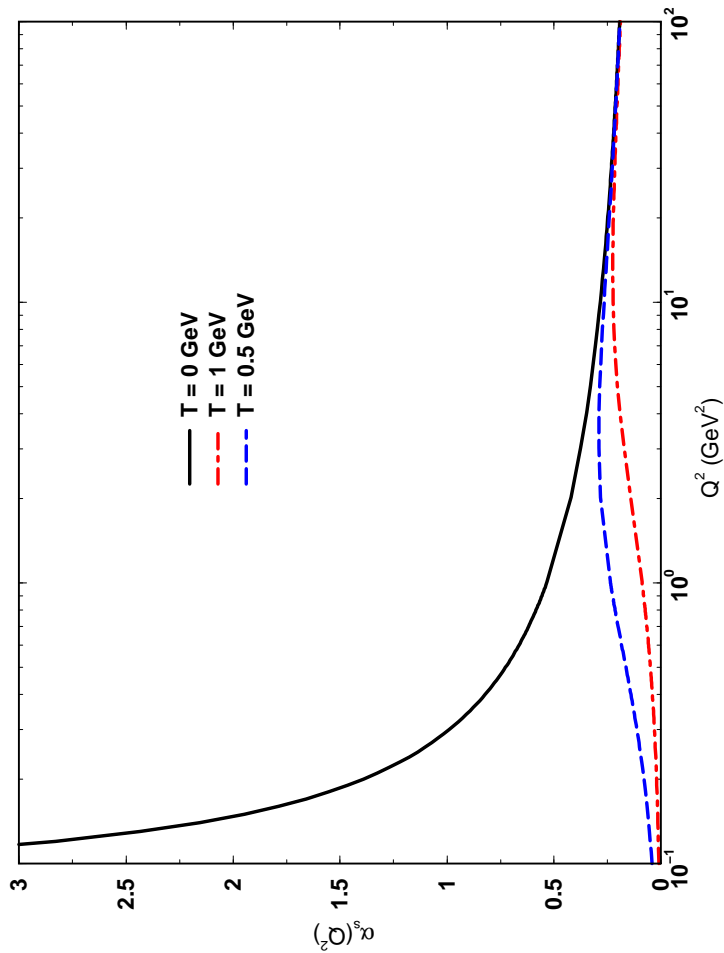}{20}{13}{2}{-13}{0.6}
\vspace{1em}
\caption{The strong coupling constant as a function of $Q^2$ calculated
for 3 differerent values of the temperature.}
\label{fig1}
\end{figure}

\begin{figure}[htb]
\graphics{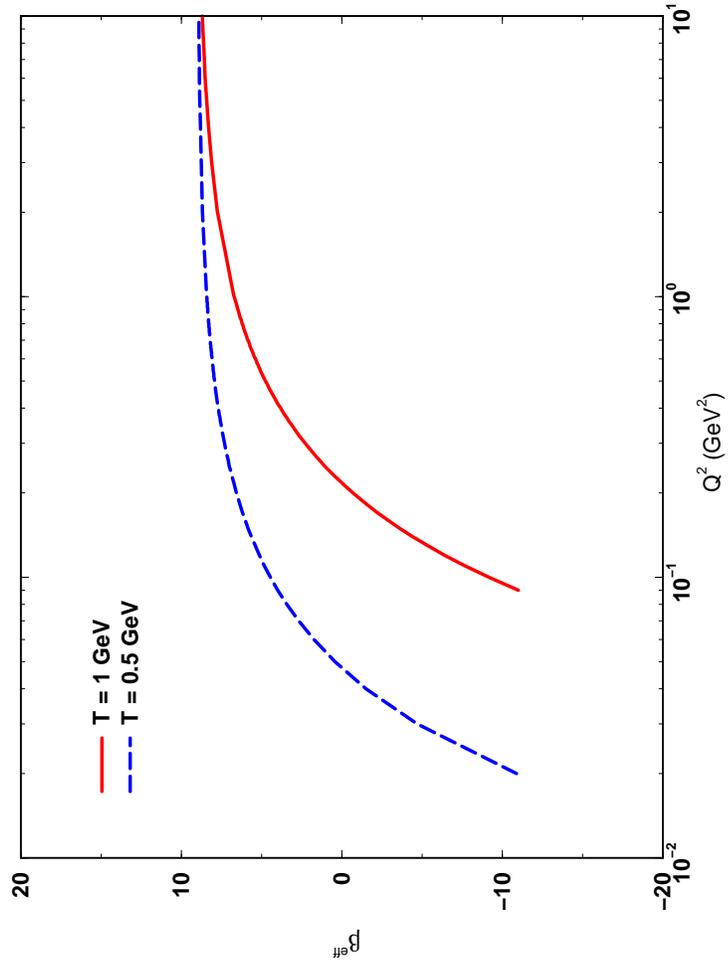}{20}{13}{2}{-13}{0.6}
\vspace{1em}
\caption{The effective beta function as a function of $Q^2$ for two 
values of the temperature. For large $Q^2$, $\beta^{eff}$ tends to 9, 
which is the value of the $\beta$ function at $T=0$ for $N_c = n_f = 0$.}
\label{fig2}
\end{figure}

In Fig. \ref{fig1} we show the behaviour of $\alpha_s$ for 3 values of the
temperature. At $T=0 \; GeV$ we have, as usual,  that the 
coupling grows rapidly for $Q^2 < 1 \; GeV^2$. However, at 
$T= 0.5$ and $ 1 \; GeV$, $\alpha_s(Q^2)$ starts to change its 
behaviour in the region around $Q^2 = 10 \; - 20\; GeV^2$. 
Instead of the 
rapid growth observed at the $T = 0$ case, for finite $T$ 
there is first an almost $Q^2$ independence of the coupling, and
then it {\it decreses} with $Q^2$. This behaviour of $\alpha_s(Q^2)$,
for finite $T$ and small $Q^2$, is that typical of an Abelian theory. 
In fact, if we define an effective $\beta$ funtion at one 
loop, 

\begin{equation}
\beta^{eff} = \frac{11}{3}N_c^{eff} - \frac{2}{3}n_f^{eff},
\label{e16}
\end{equation}
we notice that it changes sign at some small value of $Q^2$. Figure
\ref{fig2} tells us that the effective $\beta$ function changes
sign around $0.05$ and $0.2 \; GeV^2$ for $T = 0.5$ and $1 \; GeV$,
respectively. It means that at those values, the theory changes
from a non Abelian one ($\beta^{eff} > 0$) to an Abelian 
one ($\beta^{eff} < 0$), implying that at high temperature quarks
and gluons behave like electrons and photons.

Because the coupling does not grow at small $Q^2$ and high $T$,
we are allowed to use perturbation theory in this region. 
A direct consequence of this fact is that we can, 
in principle, fix $\alpha_s (Q^2\rightarrow 0, high \; T)$ using elastic 
scattering of a quark-gluon plasma by an electron beam. 

The results presented here are easily extended to
the case of a finite chemical potencial, $\mu_{cp}$. 
One just has to replace $T^2$ by $T^2 + 3 \mu_{cp}^2/\pi^2$ in the quark 
loops in the gluon self energy. The qualitative behaviour of 
$\alpha_s$ and $\beta^{eff}$ are unchanged by the introduction of
a chemical potential: the quark and gluon deconfinement at finite $T$ and
$\mu_{cp}$ still proceeds through a transition from a non Abelian
to an Abelian theory.

\acknowledgements

I would like to thank F. T. Brandt for helpful discussions on
thermal field theory, and the support of the Special Research Centre 
for the Subatomic Structure of Matter at the University of Adelaide during 
the initial stages of this work. This work was supported by
FAPESP (96/7756-6, 98/2249-4).



\end{document}